\documentclass{article}
\usepackage[utf8]{inputenc}
\usepackage{amsmath}
\usepackage{float}
\usepackage{tabularx,colortbl}
\usepackage{setspace}
\usepackage{cite,graphicx,amsmath,psfrag,bm}
\usepackage{subfigure}
\usepackage[usenames,dvipsnames]{xcolor}
\usepackage{mathabx}
\usepackage{cancel}
\usepackage{epstopdf}
\usepackage{comment}
\usepackage[process=auto]{pstool}
\usepackage[normalem]{ulem}
\graphicspath{{Figures/}}

\title{Transmissive Suppressed-Order \\Diffraction Grating (SODG)}
\author{Ashutosh Patri\textsuperscript{1}\thanks{ashutosh.patri@polymtl.ca}, Guillaume Lavigne\textsuperscript{1}, and Christophe Caloz\textsuperscript{1}}

\date{%
    \textsuperscript{1}Department of Electrical Engineering, Polytechnique Montr\'eal\\[2ex]%
    \today}

\usepackage{graphicx}

\begin{document}

\maketitle
\begin{abstract}

We present a novel type of Suppressed-Order Diffraction Grating--(SODG). An SODG is a diffraction grating whose diffraction orders have all been suppressed except one that is selected to provide electromagnetic deflection. The proposed SODG is a transmissive grating that exhibits high-efficiency refraction-like deflection for all angles, including large angles that are generally challenging to achieve, while featuring a deeply subwavelength thickness, as required in the microwave regime. We first present the design rationale and guidelines, and next demonstrate such a 10.5 GHz SODG that reaches an efficiency of $90\%$ at $70^\degree$.

\end{abstract}
\section{Introduction}

Metasurfaces, which are 2D arrays of subwavelength resonating scattering particles, possess unprecedented capabilities in manipulating electromagnetic waves, and have already led to many applications that were not achievable using conventional technologies~\cite{karim}. One of their most fundamental operations is generalized refraction~\cite{capasso}, where a phase gradient metasurface deflects the incident wave to a desired direction of space.

However, wave-deflecting metasurfaces may be challenging to realize. First, they require the design of several subwavelength scattering particles within the gradient supercell, which necessitates dense meshing and hence lengthy computation. Second, they may be difficult or impossible to fabricate depending on the resolution of the available process, due to the small features of the particles. Finally, they require more sophisticated designs than simple phase-gradient structures~\cite{capasso} to avoid supercell diffraction at large deflection angles; specifically, such metasurfaces must involve bianisotropy~\cite{guillaume}.

These issues in achieving efficient refraction-like deflection are actually specific to metasurfaces. They did not exist in the 1960s volume diffraction gratings, consisting of oblique interference fringes produced by a holographic mechanism~\cite{volume,volumerev}. However, these structures are several wavevelength thick, which hinders their application in the microwave and part of far infra-red regimes. Recently, thinner refraction-like deflection  gratings, based on resonant dielectric structures, were proposed and demonstrated in~\cite{engheta,fan}, with the latter paper referring to these structures as ``metagratings.'' At about the same time, gratings that were essentially similar, except for their reflective operation, were theoretically described in terms of surface polarizabilities in~\cite{alu}, also under the name of ``metagratings,'' and such reflective gratings were experimentally demonstrated in the microwave regime in~\cite{epstein}. The gratings reported in~\cite{alu} and~\cite{epstein} have the merit of being subwavelengthly thin, and hence perfectly suited for microwave implementations. All the gratings described in this paragraph are  diffraction gratings with mechanisms that suppress all the diffraction orders except the one that is desired. Therefore, we suggest here to call them ``Suppressed-Order Diffraction Grating (SODG)''~\footnote{Since the term ``meta'' generally refers to homogenizable structures~\cite{ari}, the terminology ``metagrating'' is not appropriate here. In contrast, SODG specifically and unambiguously describes the physics of the structure.}.

All of the deeply-subwavelengthly-thin SODGs reported so far have been designed for reflection operation~\cite{alu, epstein}. However, transmission SODGs, although harder to realize, are equally important, particularly for transmit-array applications. In this work, we propose and  demonstrate a deeply-subwavelengthly-thin \emph{transmissive} SODG, which fills up this gap.

\begin{figure}[htbp]
\centering
\includegraphics[width=0.7\columnwidth]{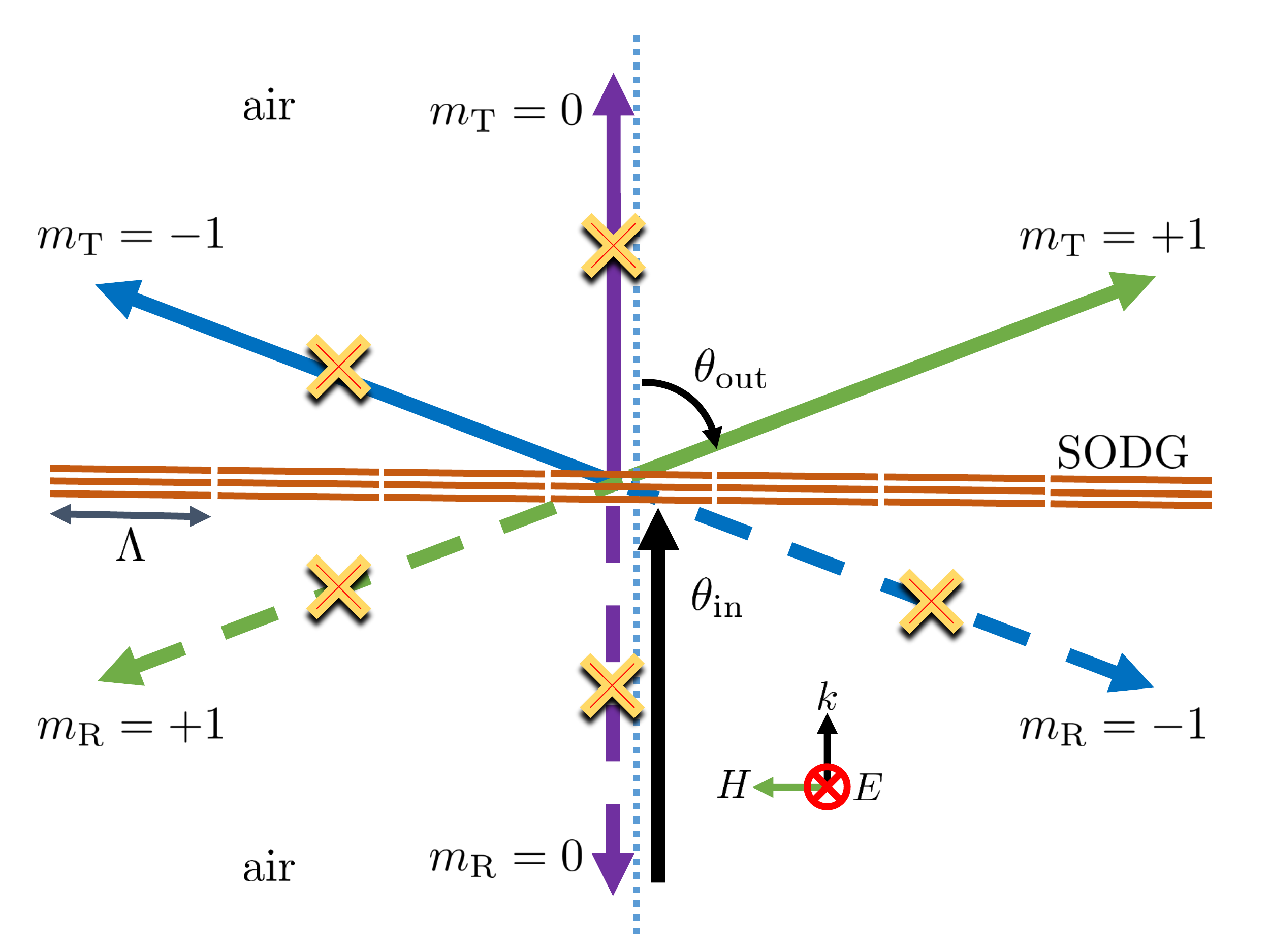}
\vspace{-3mm}
\caption{Problem of transmissive Suppressed-Order Diffraction Grating (SODG) for the efficient deflection of a normally incident beam to a direction $\theta_\text{out}$. The notation $m_\text{R}$ and $m_\text{T}$ refer to the diffraction order $m$ in the reflection and transmission, respectively. The solid black arrow represents the incident beam direction.}
\label{fig:SODG}
\end{figure} 

\section{SODG Structure and Design Rationale}

\begin{figure}[htbp]
\centering
\includegraphics[width=0.7\columnwidth, trim={0in 0in 0in 0in}]{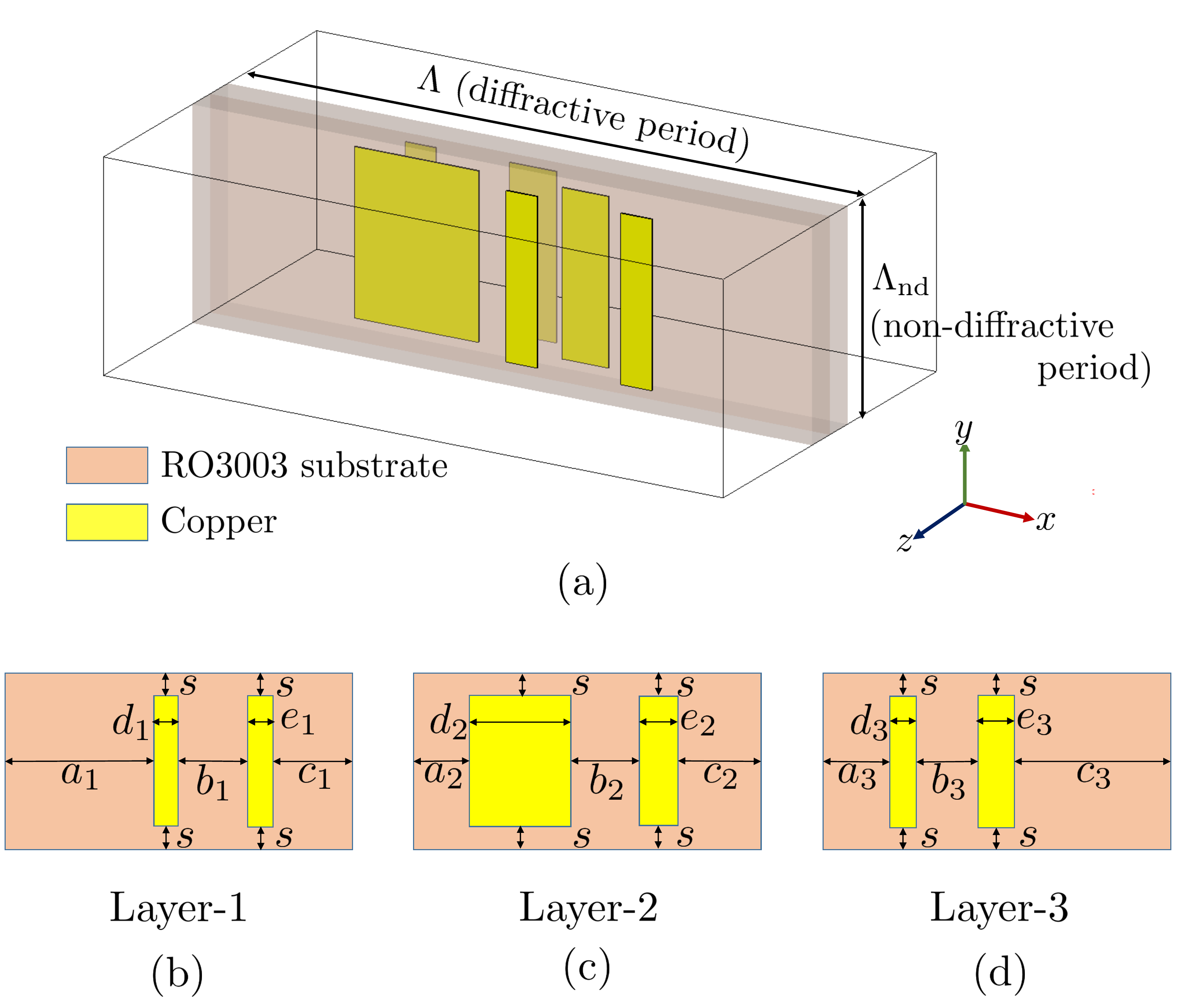}
\caption{Unit cell design parameters of a 10.5 GHz SODG deflecting a normally incident beam to the direction $\theta_\text{out}=70^\degree$. The unit cell is composed of two back-to-back 1.52~mm thick RO3003 ($\epsilon_\text{r}=3$) substrates supporting three metallization (70~$\mu$m thick copper) layers. (a)~Perspective view. (b)~First layer. (c)~Second layer. (d)~Third layer. The parameter values are, in mil: $\Lambda=1200$, $\Lambda_\text{nd}=390$, $a_{1-3}=(615.275, 273.385, 343.149)$, $b_{1-3}=(154.125, 162, 148.314)$, $c_{1-3}=(310.598, 434.614, 564.535)$, $d_{1-3}=(60, 240, 60)$, $e_{1-3}=(60, 90, 84)$, $s=42$.}
\label{fig:parameter}
\end{figure} 
Figure~\ref{fig:SODG} illustrates the transmissive SODG problem to solve. Figure~\ref{fig:parameter}(a) shows the structure we propose for that purpose, while Figs.~\ref{fig:parameter}(b)-(d) provides the corresponding design parameters for s-polarization deflection of $\theta_\text{out}=70^\degree$ from the normal at 10.5 GHz. The structure is composed of three metallization layers to ensure the possibility of completely suppressing reflection, based on the Huygens (in-plane perpendicular electric and magnetic source pair) source argument~\cite{eleftheriades}.

Our design rationale for the transmissive SODG may be divided into three steps. First, we determine the required diffractive period ($\Lambda$) for a desired set of incident and diffracted angles. Assuming air on both sides of the structure, this period follows from the grating equation as
\begin{equation}\label{EQ:Diffraction}
\Lambda=\frac{2\pi m}{k_\text{0}\sin\theta_\text{in}+k_\text{0}\sin\theta_\text{out}},
\end{equation}
where $k_\text{0}$ is the free-space wavenumber, $\theta_\text{out}$ corresponds to the (positive or negative) diffraction order $m$ selected for deflection. In order to have the smallest possible number of diffraction orders to suppress, we choose to initially have exactly 3 diffraction orders at each side of the grating, select the $m=1$ diffraction order for deflection, and identify $\theta_{m=+1}$ with $\theta_\text{out}$. This yields, from Eq.~\eqref{EQ:Diffraction}, $\Lambda=30.48$~mm.

Second, we introduce vertical asymmetry to suppress the reflective diffraction orders. Such asymmetry may be realized by using different scatterers in layers~1 and~3, in accordance with the metasurface Huygens-source principle, as shown in Fig.~\ref{fig:parameter}. 

Third, we introduce horizontal asymmetry to suppress the remaining (transmissive) diffractive orders ($m=0$ and $m=-1$). For comfortable flexibility, we realize this by using a pair of rectangular patches of different sizes in each layer and, in addition, shifting the relative positions of the pairs in different layers.

Then, the structure is optimized by full-wave simulation to minimize the amount of energy going to the undesired diffraction orders, which leads to the parameters listed in Fig.~\ref{fig:parameter} of our demonstration example.

\section{Simulation Results}

\begin{figure}[htbp]
\centering
\includegraphics[width=0.7\columnwidth]{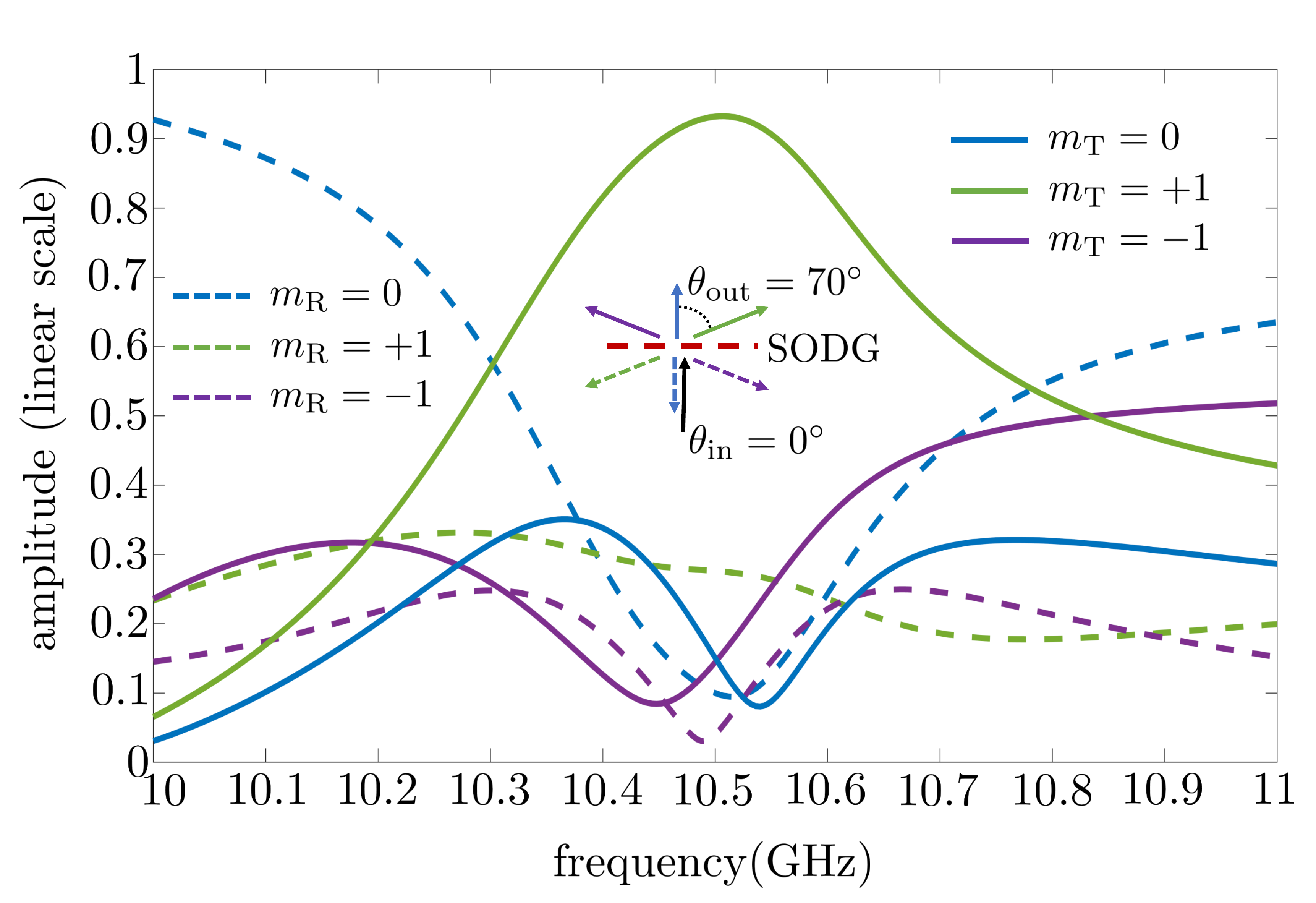}
\caption{Transmission and reflection amplitudes for the various diffraction orders, normalized to the incident beam, of the design presented in Fig.~\ref{fig:parameter}.}
\label{efficiency}
\end{figure}

Figure~\ref{efficiency} plots the full-wave simulated (FEM CST Microwave Studio) diffraction-order amplitudes versus frequency for the proposed SODG. This designs features a diffraction efficiency ($\eta=P_\text{out}/P_\text{in}=|E_{T,+1}|^2/|E_\text{inc}|^2$) of $90\%$ for the design angle of $70^\degree$ at the design frequency of $10.5$~GHz.

\section{Conclusion}

We have presented an ultra-thin ($\sim\lambda_0/10$) transmissive SODG, and demonstrated a specific design of it that exhibits an efficiency of $90\%$ at a deflection angle of $70^\degree$. This device fills up an important gap in providing a high-efficiency large-angle SODG, which are particularly needed at microwave frequencies.


\begin{thebibliography}{00}


\bibitem{karim} K. Achouri, and C. Caloz, ``Design, concepts, and applications of electromagnetic metasurfaces,'' \textit{Nanophotonics}, vol. 7, no. 6, pp. 1095--1116, 2018.

\bibitem{capasso} N. Yu, P. Genevet, M. A. Kats, F. Aeta, J. P. Tetienne, and F. Capasso,``Light propagation with phase discontinuities: generalized laws of reflection and refraction,'' \textit{Science}, vol. 7, pp. 1210713, 2011.

\bibitem{guillaume} G. Lavigne, K. Achouri, V. S. Asadchy,S. A. Tretyakov, and C. Caloz,``Susceptibility derivation and experimental demonstration of refracting metasurfaces without spurious diffraction,'' \textit{IEEE Trans. Antennas Propag.}, vol. 66, no. 3, pp. 1321--1330, 2018.

\bibitem{volume} H.  Kogelnik,  “Coupled  wave  theory  for  thick  hologram  gratings,” \textit{Bell Syst. Tech. J.}, vol. 48, no. 9, pp. 2909–-2947, 1969.

\bibitem{volumerev} S. C. Barden, J. A. Arns, W. S. Colburn, and J. B. Williams, “Volume-phase  holographic  gratings  and  the  efficiency  of  three  simple  volume-phase  holographic  gratings,” 	\textit{Publ. Astron. Soc. Pac}, vol. 112, no. 772, pp. 809--820, 2000.


\bibitem{engheta} H. Iizuka, N. Engheta, H. Fujikawa, K. Sato, and Y. Takeda, “Role of propagating  modes  in  a  double-groove grating with a +1st-order diffraction angle larger than the substrate–air critical angle,” \textit{	Opt. Lett.}, vol. 35, no. 23, pp. 3973–-3975, 2010.

\bibitem{fan} D.  Sell,  J.  Yang,  S.  Doshay,  R.  Yang,  and  J.  A.  Fan,  “Large-angle, multifunctional metagratings based on freeform multimode geometries,” \textit{Nano Lett.}, vol. 17, no. 6, pp. 3752–-3757, 2017.

\bibitem{alu} Y. Ra’di, D. L. Sounas, and A. Al\`u, ``Metagratings: beyond the limits of graded metasurfaces for wave front control,'' \textit{Phys. Rev. Lett.}, vol. 119, no. 6, pp. 1095--1116, 2017.

\bibitem{epstein} O. Rabinovich, and A. Epstein, ``Analytical Design of Printed-Circuit-Board (PCB) Metagratings for Perfect Anomalous Reflection,'' \textit{IEEE Trans. Antennas Propag.}, vol. 66, no. 8, pp. 4086--4095, 2018.

\bibitem{ari} A. Sihvola, ``Metamaterials: A Personal View,'' \textit{Radioengineering}, vol. 18, no. 2, pp. 90--94, 2009.


\bibitem{eleftheriades} M. Chen, M. Kim, A. M. Wong, and G. V. Eleftheriades, "Huygens’ metasurfaces from microwaves to optics: a review," \textit{Nanophotonics}, vol. 7, no. 6, pp. 1207--1231, 2018.

\end{thebibliography}
\end{document}